\documentclass [12pt] {article}
\usepackage{graphicx}
\usepackage{epsfig}
\begin {document}
\begin{center}

{\bf Inelastic nuclear screening for different secondaries \\
produced in $p$+Pb collisions at LHC energy} \\
\vspace{.5cm}
G.H.~Arakelyan$^1$, C.~Merino$^2$, Yu.M.~Shabelski$^3$ and A.G.~Shuvaev$^3$\\
\vspace{.5cm}

$^1$A.Alikhanyan National Scientific Laboratory
(Yerevan Physics Institute)\\
Yerevan, 0036, Armenia\\
E-mail: argev@mail.yerphi.am\\

\vspace{.2cm}

$^1$Departamento de F\'\i sica de Part\'\i culas, Facultade de F\'\i sica \\
and Instituto Galego de F\'\i sica de Altas Enerx\'\i as (IGFAE) \\
Universidade de Santiago de Compostela, Galiza, Spain \\
E-mail: merino@fpaxp1.usc.es

\vspace{.2cm}

$^{3}$ Petersburg Nuclear Physics Institute \\
NCR Kurchatov Institute \\
Gatchina, St.Petersburg 188300 Russia \\
E-mail: shabelsk@thd.pnpi.spb.ru \\
E-mail: shuvaev@thd.pnpi.spb.ru
\vskip 0.5 cm

\end{center}

\begin{abstract}
We calculate the inclusive spectra of secondaries produced in soft (minimum
bias) $p$+Pb collisions in the framework of Quark-Gluon String Model at LHC
energy, by taking into account the inelastic screening corrections
(percolation effects). The role of these effects is expected to be very large
at the very high energies, and they should decrease the spectra more than 2
times in the midrapidity region at $\sqrt s_{NN} = 5$~TeV.
The experimental data confirm such a picture, which means that
the nuclear screening effects are connected with the Pomeron interaction
rather than with the interactions
of the produced secondary particles in the final state.
\end{abstract}

\vskip 1cm


\newpage

\section{Introduction}

The investigation of soft $p$+Pb interactions
is very interesting because it can give the answer
to the problem of inelastic shadow corrections~\cite{CKTr,MPS}
for inclusive particle production.

In~\cite{CKTr,MPS} it was shown that the correct description of the
inclusive spectra of secondaries produced in d+Au collisions at
$\sqrt{s} = 200$~GeV (RHIC) requires to account for the inelastic
shadow corrections, that are probably connected with
the multipomeron interactions and that lead to the saturation of
the inclusive density of secondary hadrons in the soft (low $p_T$)
region, where the methods based on perturbative QCD cannot be used.
The effects of the inelastic shadow corrections should increase with
the initial energy. The difference in the results for the spectra obtained
from the calculations with and
without the inelastic shadow effects at LHC energies is of about a factor 2
in the midrapidity region.

The data for the inclusive densities of all charged secondaries
obtained by the ALICE ~\cite{ALICE} collaboration 
experimentally confirm the existence of these corrections~\cite{soft pPb}
at the LHC energy $\sqrt s_{NN}=5$~TeV.
In principal, two possibilities exist to explain the origin of
the inelastic nuclear screening: either it comes from the diagrams
with Pomeron interactions, or from the interactions
of the produced secondaries with another hadrons and/or Pb nucleus.
In the first case, the inelastic screening effects
should be the same for different secondaries, while for the second one
these effects should depend on the interaction cross sections of the secondaries,
so the effects should be different for the different secondaries.

In this paper we compare the experimental data for the inclusive
densities of different secondaries obtained by the  
CMS ~\cite{CMS} collaboration to the corresponding predictions of the 
Quark-Gluon String Model (QGSM)~\cite{KTM,Kaid} for $p$+Pb at $\sqrt{s}_{NN} =5$~TeV.

The QGSM quantitatively describes many features of the high energy production 
processes, including the inclusive spectra of different secondary hadrons
produced in the high energy hadron-nucleon~\cite{KaPi,Sh,ACKS,AMPS}
and hadron-nucleus collisions~\cite{KTMS,Sh1}. In the frame of the QGSM,
the hadron-nucleon interactions have already been considered at different energies,
including LHC, whereas the hadron nucleus collisions have been described
at not very high energies, where the inelastic screening corrections
are negligibly small~\cite{CKTr}.

Now, at the LHC energies the inelastic screening corrections become
large, what allows us to analyze them in more detail.

\section{Inclusive spectra of secondary hadrons in the \newline
Quark-Gluon String Model}

In order to produce quantitative predictions for the inclusive spectra of secondary hadrons,
a model for multiparticle production is needed. It is for that purpose that we have
used the QGSM~\cite{KTM,Kaid} in the numerical calculations presented below.

In the QGSM, both high energy hadron-nucleon and hadron-nucleus interactions are
treated as proceeding via the exchange of one or several Pomerons, and
all elastic and inelastic processes result from cutting through or between
Pomerons~\cite{AGK}. Each Pomeron corresponds to a cylinder diagram (see
Fig.~1a) that, when cut, produces two showers of secondaries,
as it is shown in Fig.~1b. The inclusive spectrum of secondaries is then determined
by the convolution of diquark, valence quark, and sea quark distributions
in the incident particles, $u(x,n)$, with the fragmentation functions
of quarks and diquarks into the secondary hadrons, $G(z)$.
Both functions $u(x,n)$ and $G(z)$ are
determined by the appropriate Reggeon diagrams~\cite{Kai}.

The diquark and quark distribution functions depend
on the number $n$ of cut
Pomerons in the considered diagram.
In the following calculations we use the
recipe of reference~\cite{KTMS}.

\begin{figure}[htb]
%%%\centering
%%%\vskip -2.cm
%%%\hskip 2.cm
\vskip -5.5cm
\hskip 4.cm
\includegraphics[width=.6\hsize]{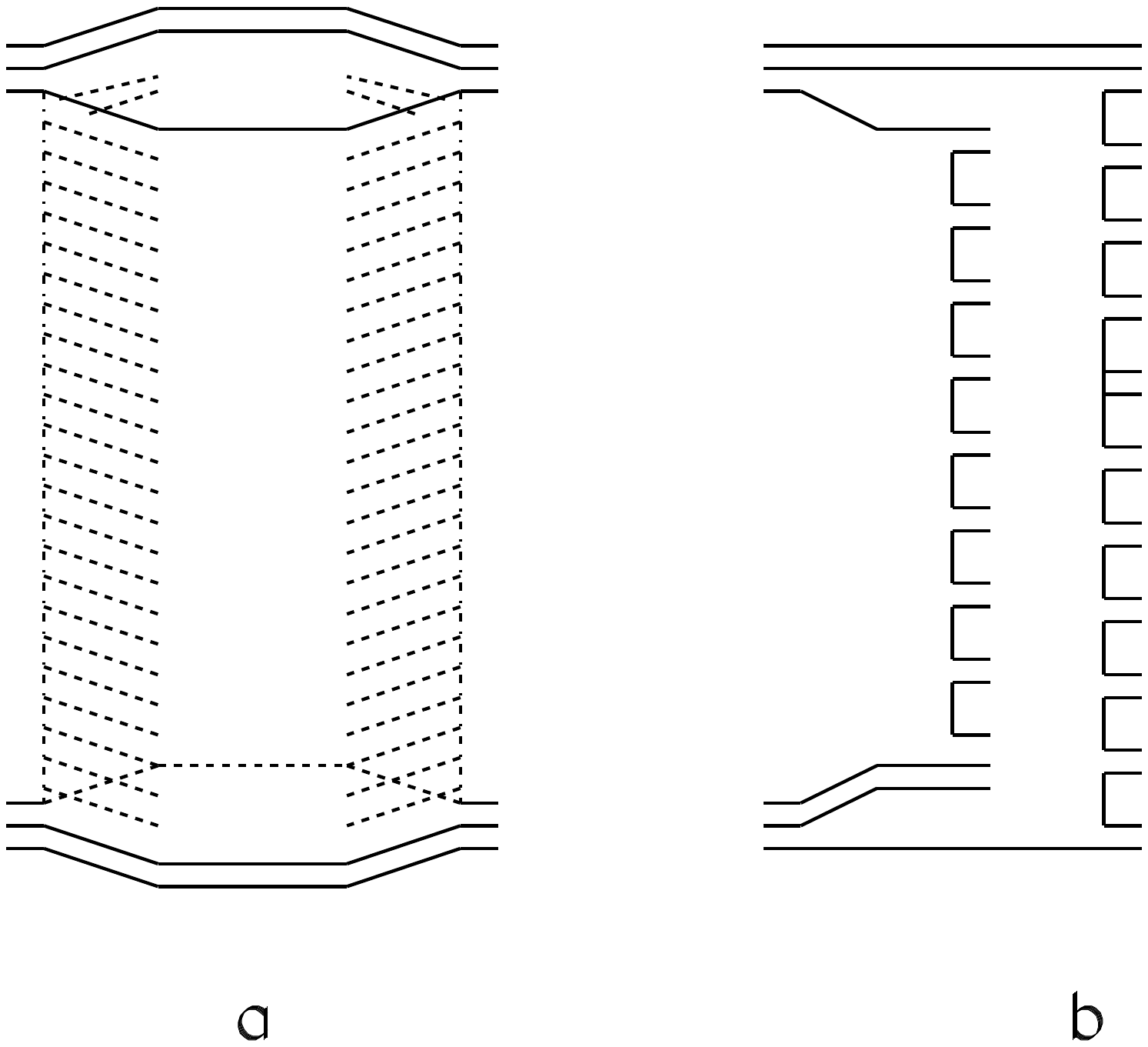}
%%%\includegraphics[width=.55\hsize]{fi00.eps}
%%%\vskip -2.5cm
\vskip -6.cm
\hskip 3.cm
\includegraphics[width=.6\hsize]{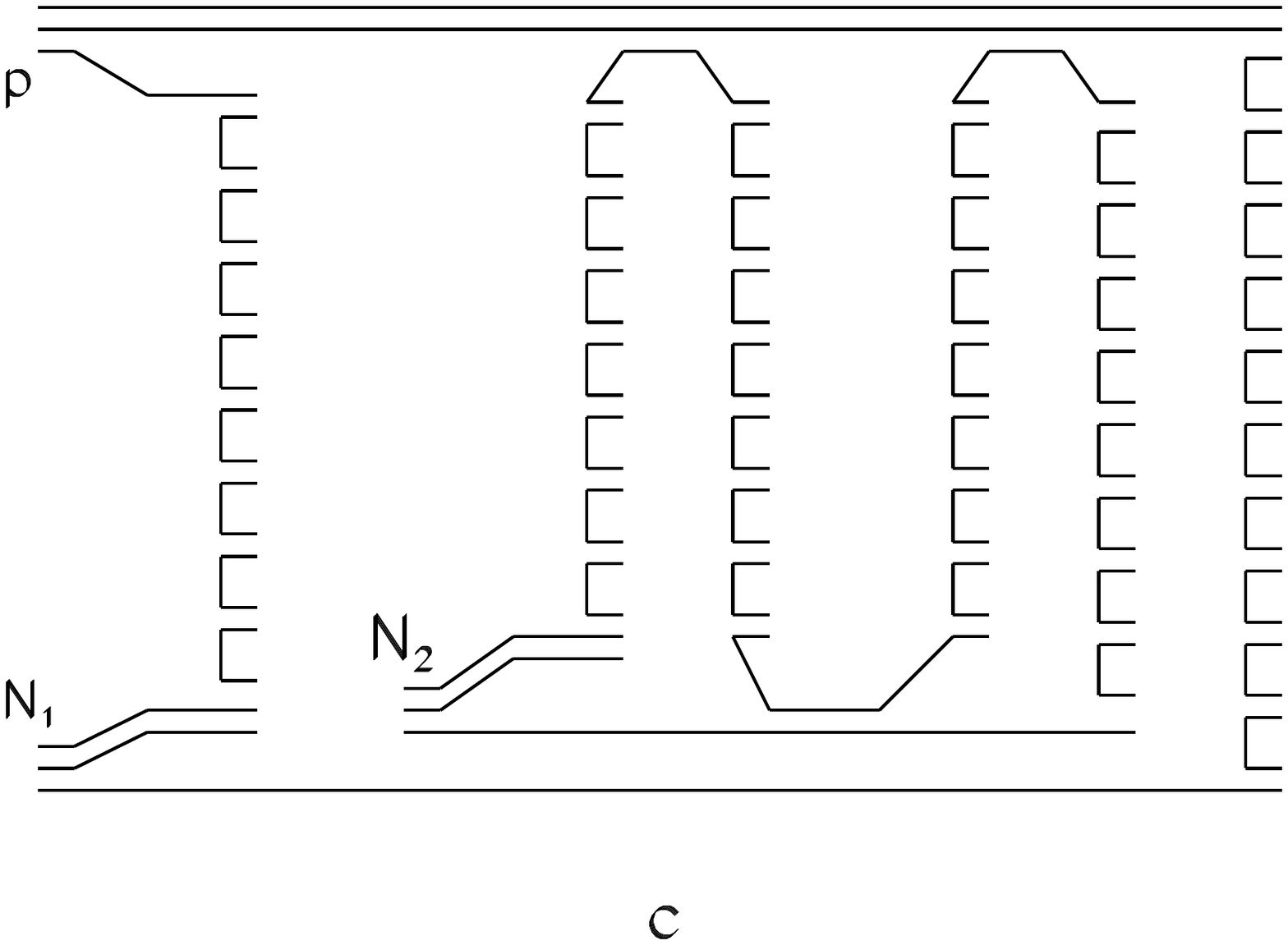}
\vskip -1.cm
\caption{\footnotesize
(a) Cylindrical diagram representing the Pomeron exchange
within the Dual Topological Unitarization (DTU) classification
(quarks are shown by solid lines);
(b) Cut of the cylindrical diagram corresponding to the single-Pomeron
exchange contribution in inelastic $pp$ scattering;
(c) Diagram corresponding to the inelastic interaction
of an incident proton with two target nucleons $N_1$ and $N_2$
in a $pA$ collision.}
\end{figure}
For the nucleon target, the inclusive density $dn/dy$ of a
secondary hadron $h$ has the form~\cite{KTM}:
\begin{equation}
\frac{dn}{dy} =  \frac{1}{\sigma_{inel}}\cdot\frac{d\sigma}{dy}
= \frac{x_E}{\sigma_{inel}}\cdot\frac{d\sigma}{dx_F}
=\sum_{n=1}^{\infty}w_{n}\cdot\phi_{n}^{h}(x)\ \ ,
\end{equation}
where the functions $\phi_{n}^{h}(x)$ determine the contribution
of diagrams with $n$ cut Pomerons, and $w_{n}$ is the probability
for this process to occur~\cite{TM}. Here we neglect the diffractive
dissociation contributions that would only be sigificant in
the fragmentation regions, i.e at large $x_F$.

For $pp$ collisions
\begin{equation}
\phi_n^{h}(x) = f_{qq}^{h}(x_{+},n) \cdot f_{q}^{h}(x_{-},n) +
f_{q}^{h}(x_{+},n) \cdot f_{qq}^{h}(x_{-},n) +
2(n-1)f_{s}^{h}(x_{+},n) \cdot f_{s}^{h}(x_{-},n)\ \  ,
\end{equation}

\begin{equation}
x_{\pm} = \frac{1}{2}[\sqrt{4m_{T}^{2}/s+x^{2}}\pm{x}]\ \ ,
\end{equation}
where $f_{qq}$, $f_{q}$, and $f_{s}$ are the contributions
of diquarks, valence quarks, and sea quarks, respectively.

These contributions are determined by the convolution
of the diquark and quark distributions with the fragmentation
functions, e.g.,
\begin{equation}
f_{q}^{h}(x_{+},n) = \int_{x_{+}}^{1}
u_{q}(x_{1},n)\cdot G_{q}^{h}(x_{+}/x_{1}) dx_{1}\ \ .
\end{equation}
The diquark and quark distributions, as well as the fragmentation
functions, are determined by Regge asymptotics~\cite{Kai}.
The numerical values of the model parameters were published
in reference~\cite{Sh}.

The probabilities $w_{n}$ in Eq.~(1) are the ratios
of the cross sections corresponding to $n$ cut Pomerons,
$\sigma^{(n)}$, to the total non-diffractive inelastic $pp$
cross section, $\sigma_{nd}$~\cite{TM}.

The contribution of multipomeron exchanges in high energy
$pp$ interactions results in a broad distribution of $w_n$ (see~\cite{soft pPb}).
In the case of interaction with a nuclear target,
the Multiple Scattering Theory (Gribov-Glauber Theory)
is used, which allows to treat the interaction with the nuclear
target as the superposition of interactions
with different numbers of target nucleons.
Let $W_{pA}(\nu)$ be the probability for
the inelastic interactions of the proton with $\nu$ nucleons of the target,
and $\sigma_{prod}^{pA}$ the total cross section of secondary
production in a p+A collision.
From the Multiple Scattering Theory, one has:
\begin{equation}
W_{pA}(\nu) = \sigma^{(\nu)}/\sigma_{prod}^{pA} \;,
\end{equation}
(see again reference~\cite{soft pPb} for the numerical examples).
Here,
\begin{equation}
\sigma^{(\nu)} = \frac1{\nu !} \int d^2b\cdot [\sigma^{pN}_{inel}\cdot T(b)]^{\nu}\cdot
e^{-\sigma^{pN}_{inel}\cdot T(b)}
\end{equation}
coincides~\cite{Sh3,BT,Weis,Jar} with the optical model
expression~\cite{TH}, and
\begin{equation}
\sigma_{prod}^{pA} = \int d^2b\cdot(1 - e^{-\sigma^{pN}_{inel}\cdot T(b)}) \:,
\end{equation}
where $T(b)$ is the profile function of the nuclear target:
\begin{equation}
T(b) = A \int^{\infty}_{-\infty} dz\cdot\rho(b,z) \:,
\end{equation}
with $\rho(r=\sqrt{b^2+z^2})$ the one-particle nuclear density.

The average value of $\nu$ has the well-known form:
\begin{equation}
\langle \nu \rangle = \frac{A\cdot\sigma^{pp}_{inel}}{\sigma^{pA}_{prod}} \;.
\end{equation}
We use the numerical values $\sigma^{pp}_{inel} = 72$~mb and $\sigma^{pPb}_{prod} = 1900$~mb
at $\sqrt{s} = 5$~TeV, so that
\begin{equation}
\label{7.8}
\langle \nu \rangle_{p+Pb}=7.8.
\end{equation}

In the calculation of the inclusive spectra of secondaries produced
in $pA$ collisions we should consider the possibility of one
or several Pomeron cuts in each of the $\nu$ blobs of the proton-nucleon
inelastic interactions.
For example, in Fig.~1c it is shown one of the diagrams contributing
to the inelastic interaction of a beam proton with two nucleons from the target.
In the blob of the proton-nucleon(1) interaction one Pomeron is cut,
and in the blob of the proton-nucleon(2) interaction two Pomerons
are cut. It is essential to take into account all the diagrams with every
possible Pomeron configuration and its permutations. The diquark
and quark distributions and the fragmentation functions here
are the same as in the case of the interaction
with one nucleon.

The process shown in Fig.~1c satisfies~\cite{Sh3,BT,Weis,Jar}
the condition that the absorptive parts of the hadron-nucleus
amplitude are determined by the combination of the absorptive parts
of the hadron-nucleon amplitudes.

\section{Inclusive spectra in p+A collisions at very high energy
and inelastic screening (percolation) effects}

The QGSM gives a reasonable description \cite{KTMS,Sh4}
of the inclusive
spectra of different secondaries produced in hadron-nucleus
collisions at energies $\sqrt{s_{NN}}$ = 14$-$30~GeV.

The situation drastically changes at RHIC energies. The spectra of
secondaries produced in $pp$ collisions can be rather well described,
but the RHIC experimental data for Au+Au collisions \cite{Phob,Phen}
give clear evidence for the inclusive density saturation effects,
which reduce the inclusive density about two times in the central
(midrapidity) region when compared to the predictions based on
the superposition picture~\cite{CMT,Sh6,AP}.
This reduction can be explained by the inelastic screening
corrections connected to multipomeron interactions~\cite{CKTr}.
The effect is very small for integrated cross sections (many of them
are determined only by geometry), but it is very important~\cite{CKTr}
for the calculations of secondary multiplicities and inclusive
densities at the high energies.

However, all estimations are model dependent. The numerical weight
of the contribution of the multipomeron diagrams is rather unclear
due to the many unknown vertices. The number of unknown parameters
can be reduced in some models, and, for example, in reference~\cite{CKTr}
the Schwimmer model~\cite{Schw} was used for the numerical estimations.

Another approaches
were used in~reference\cite{Ost}, where the phenomenological
multipomeron vertices of eikonal type were introduced
for enhancement diagram summation.

The calculations of inclusive densities and multiplicities,
both in $pp$~\cite{CP1,CP2}, and in heavy ion
collisions~\cite{CP2,CP3} (with accounting
for inelastic nuclear screening),
can be fulfilled in the percolation theory, and they result
in a good agreement with the experimental data
in a wide energy region.

The percolation model also provides a reasonable description
of the transverse momentum distribution (at low and intermediate $p_T$)
including the Cronin effect and the behavior of the baryon/meson
ratio~\cite{Dias,Paj,Paj1}.
The percolation approach assumes two or several Pomerons
to overlap in the transverse space and to fuse in a single Pomeron.
When all quark-gluon strings (cut Pomerons) are overlapping,
the inclusive density saturates, reaching its maximal value
at a given impact parameter.

In order to account for the percolation effects in the QGSM,
it is technically more simple~\cite{MPS} to consider in
the central region the maximal number of
Pomerons $n_{max}$ emitted by one nucleon.
After they are cut, these Pomerons lead to the different
final states. Then the contributions of all the diagrams
with $n \leq n_{max}$ are accounted for as at the lower energies.
The unitarity constraint
also obeys the emission of the larger number
of Pomerons $n > n_{max}$
but due to fusion in the final state (on the quark-gluon
string stage) the cut
of $n > n_{max}$ Pomerons results in the same final
state as the cut of $n_{max}$ Pomerons.

By doing this, all model calculations become rather simple
and very similar to those in the percolation approach.
The QGSM fragmentation formalism allows one to
calculate the spectra of different secondaries integrated
over $p_T$ as functions of initial energies, rapidity, and $x_F$.
In this scenario we obtain a reasonable agreement with
the experimental data
on the inclusive spectra of secondaries at RHIC energy
(see~\cite{MPS} with $n_{max} = 13$).

It has been shown in~\cite{JDDCP} that the number of strings
for the secondary production
should increase with the initial
energy even when the percolation effects are included.
Thus, in the following calculations we use the value
$n_{max} = 21$
at the LHC energy $\sqrt{s} = 5~TeV$, that can be regarded
as the normalization of all the charged
secondaries multiplicities in the midrapidity region
to the ALICE data \cite{ALICE}.
The predictive power of our calculation
applies for different sorts of secondaries in midrapidity region.
If the inelastic nuclear screening comes mainly from
the Pomeron interactions, as it was discussed above,
the screening effects would be the same for all the secondaries.
On the other hand, if the final state absorption of the produced
particles are important, nuclear screening effects
would be different for different secondaries, i.e. for kaons
and antibaryons.

In the following calculations, one additional effect is also
taken into account, namely the transfer of the baryon charge to
large distances in rapidity space through the string junction
effect~\cite{ACKS,MRS}.
% \cite{ACKS,Artr,IOT,RV,Khar,MRS,SJ2}.
This transfer leads to an asymmetry in the production
of baryons and antibaryons in the central region
that is non-zero even at LHC energies.
% ~\cite{MPRS}.
In the calculation of these effects,
the following values have been chosen for the model parameters~\cite{MRS}:
\begin{equation}
\alpha_{SJ}\, =\, 0.5\;\; {\rm and} \quad \varepsilon\, =\, 0.0757\,.
\end{equation}

\section{Rapidity spectra of different secondaries at LHC energies}

To compare the calculated effect of nuclear
screening with the experimental data, the adequate description
of the secondary production on nucleon, as well as on nuclear targets
is needed. First, we present the QGSM description of $\pi^\pm$,
$K^\pm$, $p$, and $\overline{p}$ productions in $pp$ collisions at LHC energies,
and then we compare the results of our calculations with the experimental data by
the CMS Collaboration~\cite{CMS,CMSpp}
and by the ALICE Collaboration~\cite{ALICEpp09t,ALICEpp276t,ALICEpp7t}, as it is shown 
in Fig~2, where, following the analysis published by the ALICE Collaboration~\cite{ALICEpp7t},
the productions of average $\pi$, $K$, and  $p\overline{p}$ are presented. 

As it can be seen in Fig.~2, the experimental data by the ALICE Colllaboration are 
approximately 20$-$30\% lower than those published by the CMS Collaboration.  
%Taking into account some normalisations discrepane between CMS and ALICE 
%collaborations experimental data, 
The agreement of QGSM calculations with both sets of experimental data looks 
rather reasonable, given that the accuracy of our calculations
is estimated to be on the level of a 10$-$15\% theoretical uncertainty.
\begin{figure}[htb]
%%%\centering
%%%\vskip -0.5cm
\vskip -5.cm
\hskip 2.5cm
\includegraphics[width=.7\hsize]{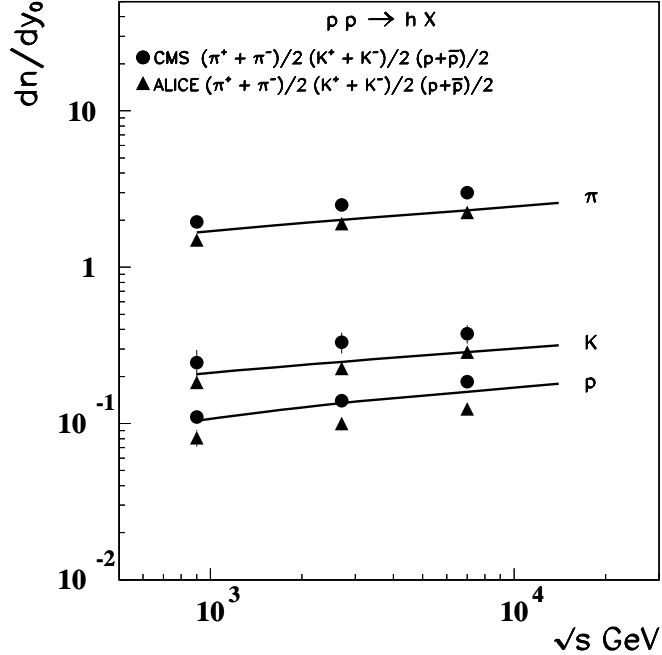}
\vskip -.5cm
\caption{\footnotesize
The energy dependence of the rapidity density $dn/dy$ at $y=0$ of average pions,
kaons and protons/antiprotons production in $pp$ collisions.
The experimental data are by the CMS Collaboration~\cite{CMSpp,CMS} 
and by the ALICE collaboration~\cite{ALICEpp09t,ALICEpp276t,ALICEpp7t}. 
The theoretical curves represent the result of the corresponding QGSM calculations.}
\end{figure}

Now, let us consider the normalization of the QGSM calculations
for the case of nuclear targets
to the experimental point by the ALICE Collaboration~\cite{ALICE},
$dn_{ch}/d\eta = 16.81 \pm 0.71$
at $\sqrt{s_{NN}} = 5$~TeV. 
The agreement is reached at $n_{max} = 21$, here we have
$dn/d\eta$ for $|\eta|<2$.
Later, the experimental value $dn/dy(|y|\leq 1) =19.1\pm 0.2$ has been
published by the CMS Collaboration~\cite{CMS}, while the QGSM calculation gives
$dn/dy(|y|\leq 1)=19.11$ with $n_{max} =21$,
so we can use this $n_{max}$ value in our analysis.

The experimental data for $p$+Pb collisions by the CMS Collaboration on the
inclusive densities of different secondaries, $\pi^\pm$,
$K^\pm$, $p$, and $\overline{p}$~\cite{CMS} are presented in
Table~1, where they are compared with our QGSM predictions.
The agreement for every secondary particles is good, what it
means that the experimental nuclear shadowing factor
is the same for different secondaries, as it is assumed in our
calculations.

Also in Table~1, we present the QGSM predictions for
the $pp$ collisions at the same energy.
The ratios of particle yields in $p$+Pb and $pp$ collisions
are equal to 3.6$-$3.7, i.e they are two times smaller than
the values of $\nu_{p+Pb}$ in Eq.~\ref{7.8}.
In the absence of inelastic nuclear screening, the ratio
$r =pPb/pp$ in the midrapidity region should be equal
to $\nu_{p+Pb}$~\cite{Sh3,BT,Weis,Jar},
that is, to the average number of the inelastic collisions
of the incident proton in the target nucleus.
Thus, we can see that the inelastic nuclear screening factor
is little larger than 2, and it is practically the same
for all considered secondaries.

\begin{center}
\vskip 5pt
\begin{tabular}{|c||c||c|c|c|} \hline
particles &CMS Collaboration & \multicolumn{3}{c|}{QGSM} \\
\cline{3-5}
% particles &CMS Collaboration & QGSM  & QGSM &  \\
        & $dn/dy$, $|y|\leq 1$~\cite{CMS} &       $p$+Pb &  $pp$ & $r$\\
\hline
\hline
$\pi^+$  & $8.074 \pm 0.087$ & 8.103 & 2.190 & 3.70\\ \hline

$\pi^- $ & $7.971 \pm 0.079 $ & 7.923 & 2.147 & 3.69  \\ \hline

$K^+$ & $1.071 \pm 0.069 $ &1.006 &0.273 & 3.69 \\ \hline

$K^-$ & $0.984 \pm 0.047 $ &0.996 &0.271 & 3.66 \\ \hline

$p$ & $0.510 \pm 0.018 $ &0.545 &0.150 & 3.63 \\ \hline

$\bar p$ & $0.494 \pm 0.017 $ &0.536 &0.148 & 3.62  \\ \hline

\hline
\end{tabular}
\end{center}
Table~1: {\footnotesize Experimental data on $dn/dy$, $|y|\leq 1$ by the
CMS Collaboration~\cite{CMS} of charged pions, kaons,
$p$, and $\overline{p}$ production in central $p$+Pb collisions
at $\sqrt{s_{NN}}$ = 5~TeV, together with
the corresponding QGSM results. The parameter
$r$ is the ratio of the particle yields in $p$+Pb and $pp$
reactions. The predictions for $pp$ collisions are also given.}

Our predictions for hyperon and antihyperons production
in $pp$ and $p$+Pb collisions at the same energy $\sqrt s=5$~Tev
are presented in Table~2.

The ratios of the inclusive densities of all secondary
hyperons and antihyperons
produced on Pb and hydrogen targets are practically the same
as for secondary mesons production, with a $\sim 5\%$ accuracy
(see tables~1 and~2).
If our predictions will be experimentally confirmed, that would
indicate that the main contribution to the processes of hyperon
and meson production has a similar nature.

\begin{center}
\vskip 5pt
\begin{tabular}{|c||c|c|c|} \hline
particles   & $p$+Pb $dn/dy_{y=0}$& pp $dn/dy_{y=0}$& $r$ \\
\hline
\hline
$\Lambda $  &  0.307 & 0.0843 &  3.64\\ \hline

$\bar \Lambda $ &  0.303 & 0.0827 & 3.66\\ \hline

$ \Xi^- $ & 0.0250 &0.00676 & 3.70 \\ \hline

$\bar \Xi^+ $ & 0.0248 &0.00669 & 3.70 \\ \hline

$\Omega^-$ & 0.00143 &0.000401 & 3.57 \\ \hline

$\bar \Omega^+ $ & 0.00142 &0.000397 & 3.58 \\ \hline
\end{tabular}
\end{center}
Table~2: {\footnotesize The QGSM prediction for the densities of hyperons and antihyperons
production $dn/dy_{y=0}$, in $p$+Pb and pp collisions at $\sqrt s$ = 5.02 TeV.}

\section{Conclusion}

It is seen that the inelastic nuclear screening corrections
at LHC energies are really large. All the ratios of inclusive
densities of the secondaries produced in lead and hydrogen targets
take values $dn/dy$, $|y|\leq 1$ = 3.6$-$3.7 (see Table~1), instead
of the values $dn/dy$, $|y|\leq 1$ = 7.5$-$8.0 that one would expect
in the absence of this effect (see Eq.~\ref{7.8}).

In our approach it is naturally explained that the nuclear screening
effects do not depend on the secondary produce particle, and 
they are practically the same (within our theoretical accuracy)
for $\pi^\pm$, $K^\pm$, $p$, and $\overline{p}$ production.
This will be checked, and we are confident that confirmed,
when the nuclear screening effects for the cases
of hyperon production will be measured (see our predictions in Table~2).

%%%In our approach it is naturally explained that the nuclear screening effects
%%%do not depend on the secondary produce particle, and they
%%%are practically the same (within our theoretical accuracy)
%%%for $\pi^\pm$, $K^\pm$, $p$, and $\overline{p}$ production, this fact
%%%being further checked and confirmed in
%%%the case of hyperon production (see our predictions in Table~2).
\vskip 0.4cm

{\bf Acknowledgements}

We thank C.~Pajares for his valuable comments.

This work has been supported by Russian RSCF grant No. 14-22-00281,
by the State Committee of Science of the Republic of Armenia, Grant-15T-1C223,
by Ministerio de Ciencia e Innovaci\'on of Spain under project
FPA2014-58293-C2-1-P, and the Spanish Consolider-Ingenio 2010 Programme CPAN (CSD2007-00042),
and by Xunta de Galicia, Spain (2011/PC043).

\end {document}